\documentclass[pdflatex,sn-mathphys-ay]{sn-jnl}
\usepackage[utf8]{inputenc}
\usepackage{geometry}
\usepackage{setspace}
\usepackage{natbib}
\usepackage{bm}
\usepackage{amsmath}
\usepackage{amssymb}
\usepackage{graphicx}
\usepackage{booktabs}
\usepackage{threeparttable}
\usepackage{tabularx}
\usepackage{comment}

\usepackage{adjustbox}
\usepackage{amsthm}

\newtheorem*{remark}{Remark}
\begin{document}
\title[Weighted Brier Score - an Overall Summary Measure for Risk Prediction Models with Clinical Utility Consideration]{Weighted Brier Score - an Overall Summary Measure for Risk Prediction Models with Clinical Utility Consideration}

\author*[1]{\fnm{Kehao} \sur{Zhu}}\email{zhukehao@gmail.com}

\author[2]{\fnm{Yingye} \sur{Zheng}}\email{yzheng@fredhutch.org}

\author[1]{\fnm{Kwun Chuen Gary} \sur{Chan}}\email{kcgchan@uw.edu}

\affil*[1]{\orgdiv{Department of Biostatistics}, \orgname{University of Washington}, \city{Seattle}, \postcode{98195}, \state{WA}, \country{USA}}

\affil[2]{\orgdiv{Public Health Sciences Division}, \orgname{Fred Hutchinson Cancer Center}, \city{Seattle}, \postcode{98019}, \state{WA}, \country{USA}}

\abstract{As advancements in novel biomarker-based algorithms and models accelerate their use in disease risk prediction, it is crucial to evaluate these models within the context of their intended clinical application. Prediction models output the absolute risk of disease; subsequently, patient counseling and shared decision-making are based on the estimated individual risk and cost-benefit assessment. The overall impact of the application is referred to as clinical utility, which received significant attention and desire to incorporate into model assessment lately. The classic Brier score is a popular measure of prediction accuracy; however, it is insufficient for effectively assessing clinical utility. To address this limitation, we propose a class of weighted Brier scores that aligns with the decision-theoretic framework of clinical utility. Additionally, we decompose the weighted Brier score into discrimination and calibration components, and we link the weighted Brier score to the $H$ measure, which has been proposed as an alternative to the area under the receiver operating characteristic curve. This theoretical link to the $H$ measure further supports our weighting method and underscores the essential elements of discrimination and calibration in risk prediction evaluation. The practical use of the weighted Brier score as an overall summary is demonstrated using data from a prostate cancer study.}


\maketitle

\section{Introduction}
Risk prediction models are widely developed and used in medicine. Various assessment measures have been proposed and studied to evaluate the performance of a risk model or compare rival risk models. Two general aspects of the assessment are discrimination and calibration \citep{steyerberg2010assessing,pfeiffer2017absolute}. Discrimination measures the degree of separation between the predicted risks for cases and controls, and it is usually assessed by the area under the receiver operating characteristic curve (AUC), a common metric reported in the literature. In contrast, the assessment of calibration receives less attention \citep{van2019calibration}. Calibration refers to how well the predicted risks agree with the observed event rates in certain subgroups. For practical assessment, the subgroups are often defined by deciles of predicted risks.

The Brier score, mean squared error of predicted risks, is an overall measure of accuracy, which comprises both discrimination and calibration \citep{brier1950verification,pfeiffer2017absolute}. When comparing two risk prediction models, one may outperform the other in discrimination but worse in calibration. In such cases, the Brier score is particularly useful as a single measure that evaluates both discrimination and calibration. Hence, the Brier score and its scaled version, i.e., the index of prediction accuracy (IPA), have been promoted recently in the risk prediction literature as an overall measure for model comparison \citep{rufibach2010use,hilden2014note,kattan2018index}.  To illustrate this point, we provide a numerical example in Section \ref{simulation} for model comparison where one model has a higher AUC, but another model is substantially better calibrated.

On the other hand, the Brier score has been viewed as unsuitable to evaluate the clinical utility of a risk prediction model \citep{pfeiffer2017absolute,assel2017brier}. In this context, the clinical utility refers to the overall impact of implementing these prediction models within their intended applications. To incorporate clinical utility into the assessment of the risk prediction models, \cite{gail2005criteria} discussed a decision-theoretic framework that assigns different costs to decisions made based on a risk prediction model with an optimal risk cutoff corresponding to the cost trade-offs specific to a particular application. In line with this decision-theoretic framework, \cite{vickers2006decision} proposed the net benefit at an optimal risk cutoff $c$.  The simplicity of the net benefit makes the methodology popular in recent medical literature to measure the clinical utility. Efficient study designs and formal statistical inference have been carefully developed to meet the exploding popularity \citep{marsh2020statistical,pfeiffer2020estimating,sande2020statistical}. However, net benefit is a function of cutoff $c$ and it is often challenging to specify a fixed $c$ to reflect a constant cost ratio in a population.  The decision curve analysis examines the net benefit at a range of $c$ and can be used as a sensitivity analysis, but it may be oversimplified to assume that the population shares the exact same costs at each specific $c$ on the decision curve \citep{kerr2016assessing}.

In the presence of many summary measures for describing the risk prediction model performance, it is helpful to consider the concept of proper scoring rules, which draws the most attention from the weather forecast and the machine learning communities \citep{gneiting2007strictly,brocker2009reliability,reid2010composite} and is relatively less known to the medical risk prediction community \citep{hilden2014note,pepe2015net}. An assessment measure is a \textit{proper} scoring rule if the scoring rule is optimized at the true risk. The scoring rule is \textit{strictly proper} if the optimizer is unique. 

Inspired by the study of proper scoring rules and various caveats of existing assessment measures, particularly the lack of clinical utility consideration, we study a weighted Brier score, which incorporates calibration, discrimination, as well as clinical utility. In contrast, \cite{gerds2008performance} proposed a different version of the weighted Brier score with a method of weighting not in line with the decision-theoretic framework \citep{gail2005criteria}; hence, it is hard to specify or interpret those weights. Instead, we propose a weighted Brier score by appealing to the results from the study of proper scoring rules, in particular, the integral representation \citep{schervish1989general}. This systematic way of weighting is coherent with the decision-theoretic framework. Furthermore, we decompose the weighted Brier score into calibration, discrimination, and uncertainty components. After studying the decomposition carefully, we also make the connection with the $H$ measure, an alternative to the AUC that can incorporate the utility for discrimination assessment of a general real-valued prediction score \citep{hand2009measuring}, which has been cited more than 1000 times. The similarity between the weighted Brier score and the $H$ measure demonstrates the validity and potential wide appreciation of the proposed weighting method. Their difference highlights the calibration aspect of risk prediction, which is crucial for patient counseling and decision-making.

The following is an outline of this paper. We first study a connection between the net benefit and the Brier score using two key results: the decision-theoretic framework for clinical utility and the integral representation in Section \ref{risk framework}. After recognizing the connection, we formulate and study the weighted Brier score in Section \ref{wBS}. In Section \ref{simulation}, we illustrate the merits of this new measure with some numerical examples in simulated data and a real data set. 

\section{Clinical utility in risk-based decision framework} \label{risk framework} 
Let $Y=1$ and $Y=0$ denote the binary outcomes of individuals with disease (cases) and without disease (controls), respectively, which is only observable in the future or costly to observe (e.g., via invasive biopsy); $\pi=P(Y=1)$ is the disease prevalence. Let $r(\bm{x}):=P(Y=1|\bm{X}=\bm{x})$ be the probability or the risk of being disease conditional on $\bm{X}=\bm{x}$, a set of predictors. The goal is the evaluation of a given risk model, $\hat{r}(\bm{x}) \in (0,1)$, which is pre-trained from an external training dataset.   The evaluation uses an independent validation dataset with $n$ copies of independent and identically distributed (i.i.d.) prediction and outcome pairs, ($\hat{r}_i$,$Y_i$), where $\hat{r}_i:=\hat{r}(\bm{X}_i)$ and $i=1,...,n$. 

To facilitate the evaluation, we define loss functions $\ell(\hat{r},Y) \in [0,\infty)$, which assign a non-negative score to the prediction, $\hat{r}$, paired with the observed outcome $Y$. For example, the square loss is $(\hat{r}-Y)^2$. To evaluate the predicted risk in a population, the parameter of interest is $E_{\bm{X},Y}[\ell(\hat{r}(\bm{X}),Y)]$, the expected value of a loss function marginalizing over both the predictors and the observed outcome. In a sample, we use the empirical average for the expectation. The Brier score, as commonly defined \citep{brier1950verification,hilden2014note,pfeiffer2017absolute}, is the empirical average of the squared loss: $\frac{1}{n}\sum^{n}_{i=1} (\hat{r}_i-Y_i)^2$. 

We focus on evaluating a risk model in light of its application in a target population and the overall impact of such application is termed clinical utility. In a risk-based decision framework, the application is whether to recommend an intervention based on risk. For example, if an individual has a high risk of cardiovascular disease in the next ten years, we recommend taking statin; if an individual has a high risk of progressive cancer, we recommend a biopsy. Hereafter, we refer to all interventions as treatment. In particular, we recommend treating a patient if $\hat{r}$ is greater than a cutoff $c$ (i.e., $I[\hat{r}>c]$, where $I[.]$ is the indicator function and 1 means recommending a treatment). Given this cutoff $c$, we define a cost-weighted misclassification loss \citep{buja2005loss,reid2010composite} to balance the trade-off of false positives and false negatives:
\[\ell(\hat{r},Y;c):=cI[\hat{r}>c](1-Y)+(1-c)I[\hat{r}\leq c]Y,
\]
and the corresponding cost-weighted misclassification error $L(c)$ is
\[
L(c):=E_{\bm{X},Y}[\ell(\hat{r}(\bm{X}),Y;c)]=cP(\hat{r}>c,Y=0)+(1-c)P(\hat{r}\leq c,Y=1) \ .
\]

%$\ell(\hat{r},Y;c)$ is weighted by the cost in the following sense. Suppose $C_1$ is the cost of not treating a case (false negative) and $C_0$ is the cost of treating a control (false positive), the total cost of a risk-based decision is 

%\begin{equation}\label{tc}
%    C_0P(\hat{r}>c,Y=0)+C_1P(\hat{r}\leq c,Y=1).
%\end{equation}

%It is known that a rational choice of $c$ to minimize the total cost is $C_0/(C_0+C_1)$ \citep{pauker1975therapeutic,elkan2001foundations,gail2005criteria}  \gary{Why do you use $(1+C_1/C_0)^{-1}$?  It is too convoluted and readers will not like it.} . 
%In other words, the choice of $c$ reflects the relative cost of false negatives and false positives. We may standardize the total cost, equation (\ref{tc}), by $1/(C_0+C_1)$ and with the rational choice of $c$ \gary{I have a hard time understanding the logic from just reading it, since $L(c)$ is defined for all $c$, there is no $C_0$ and $C_1$, and what is $argmin_c L(c)$ because that is what matters?}, we arrive at the cost-weighted misclassification error $L(c)$: 
%\begin{equation*}
%    \begin{align*}
%    &\frac{C_0}{C_0+C_1}P(\hat{r}>c,Y=0)+\frac{C_1}{C_0+C_1}P(\hat{r}\leq c,Y=1)\\
%    =&cP(\hat{r}>c,Y=0)+(1-c)P(\hat{r}\leq c,Y=1)=:L(c)
%    \end{align*}
%\end{equation*}
%Note that $L(c)=E[\ell(\hat{r},Y;c)]$. 

Due to the construction within the same risk-based decision framework, $L(c)$ is closely linked to the notion of net benefit measures, which is an increasingly popular measure to summarize the clinical utility \citep{vickers2006decision,steyerberg2010assessing,marsh2020statistical}. In Appendix \ref{app: risk framework}, we provide a review of the shared constructions of $L(c)$ and the net benefit measures. 

\cite{assel2017brier} used simulation studies to demonstrate the advantages of net benefit measures over the Brier score for evaluating clinical utility. While they suggested a weighted Brier score, they did not provide specific details on how to implement it. Here, we first elucidate the inability of the Brier score for evaluating clinical utility by noting the following fact: 

\begin{equation} \label{bs}
    E[BS]:=E_{\bm{X},Y}[\frac{1}{2}(\hat{r}(\bm{X})-Y)^2]=\int_{0}^{1}L(c)dc,
\end{equation}
where $BS:=\frac{1}{2n} \sum_{i=1}^{n}(\hat{r}_i-Y_i)^2$. Note that for equation (\ref{bs}), we include a constant $1/2$ in defining $BS$, which is irrelevant for comparing risk models. Equation (\ref{bs}) says the expected value of Brier score is the average of $L(c)$ with the uniform weight over $(0,1)$, where $L(c)$ measures the clinical utility of a risk prediction model at a specific cutoff $c$. The integral representation of the Brier score is a known result in the literature on proper scoring rules \citep{schervish1989general,gneiting2007strictly,reid2011information}. However, its application to the evaluation of the clinical utility of risk models has not, to the best of our knowledge, been previously discussed.

In many clinical applications, although it can be hard to pinpoint a specific risk cutoff $c$, there might be a general knowledge of the distribution or possible ranges of cost ratios. Thus, to reflect this knowledge, it is natural to consider a loss function with an asymmetric weight function; we call the resulting summary measure the weighted Brier score and study it in detail in the next section.

\section{Weighted Brier score} \label{wBS}
In this section, we first define the weighted Brier score and provide interpretations of the weight function. We then examine the decomposition of the weighted Brier score and make connections to some related quantities. 

\subsection{Definition, interpretations and estimation} \label{wBS def}

We consider a weight function, $w(c)>0$, as a probability density function (PDF) with support over $(0,1)$, which reflects a priori knowledge about the cost ratio in a particular clinical application of a risk model. The weighted Brier score is defined as:

\begin{equation}
    BS_w:=\frac{1}{n}\sum_{i=1}^{n} \ell_w(\hat{r}_i,Y_i),
\end{equation}
where $\ell_w(\hat{r}_i,Y_i):=\int_{0}^1\ell(\hat{r}_i,Y_i;c)w(c)dc$; the expected value of the weighted Brier score is 
\begin{equation}
E[BS_w]=E_{\bm{X},Y}[\ell_w(\hat{r}(\bm{X}),Y)]=\int_{0}^{1}L(c)w(c)dc.
\end{equation} 

We illustrate the interpretations of the weight functions with Beta distributions. Figure 1 displays the density functions of several Beta distributions. Beta distributions have been used as informative priors in clinical trials to flexibly incorporate expert opinions; see \cite{wu2008elicitation} for the elicitation of model parameters when experts cannot specify the quantities with precision. As noted earlier, the Brier score uses the uniform distribution, $Beta(1,1)$, as the weight function. If we believe that for the majority of the population, the cost of false negatives, denoted as $C_1$, is generally larger than the cost of false positives, denoted as $C_0$, we would want to favor small cutoff values so that the chance of having false negatives will be kept low.  In such a case, we may choose $Beta(3,15)$ or $Beta(2,8)$ but not $Beta(4,3)$. The two distributions $Beta(3,15)$ and $Beta(2,8)$ share a common mode of 1/8, representing two populations;  both populations have a significant concentration of individuals with cost ratios close to $C_1/C_0=7$. See Appendix \ref{app: risk framework} for a more detailed discussion of the relationship between the cost ratio and the cutoff. Two possible interpretations regard the difference between $Beta(3,15)$ and $Beta(2,8)$. First, we may believe that $Beta(3,15)$ better characterizes a more homogeneous population with less variable cost ratios. For example, in a population of cancer patients facing a decision to receive an aggressive surgery treatment, relatively young patients may have a lower risk cutoff of cancer progression to receive surgery than old patients; a population of patients with a similar age is expected to be more homogeneous in terms of the cost ratio. As an alternative interpretation, we may choose $Beta(2,8)$ because we are not sure about the cost ratios in the population and are reluctant to put too much weight on a certain region. If we are certain that everyone in the population has the same cost ratio (e.g., $C_1/C_0=7$), we may choose a point mass function at 0.125 as $w(c)$, which results in $L(c=0.125)$ or corresponding net benefit measures to assess the clinical utility.  The exact choice of the weight function is not necessary, and we will show using a real data example in Section 5.2 that $Beta(3,15)$ and $Beta(2,8)$ give consistent rankings different from $Beta(4,3)$ and $Beta(1,1)$, which illustrate a specification of the weight function broadly consistent with clinical utility would be sufficient in practice. 
\begin{figure}[h]
\centerline{\includegraphics[scale=0.60]{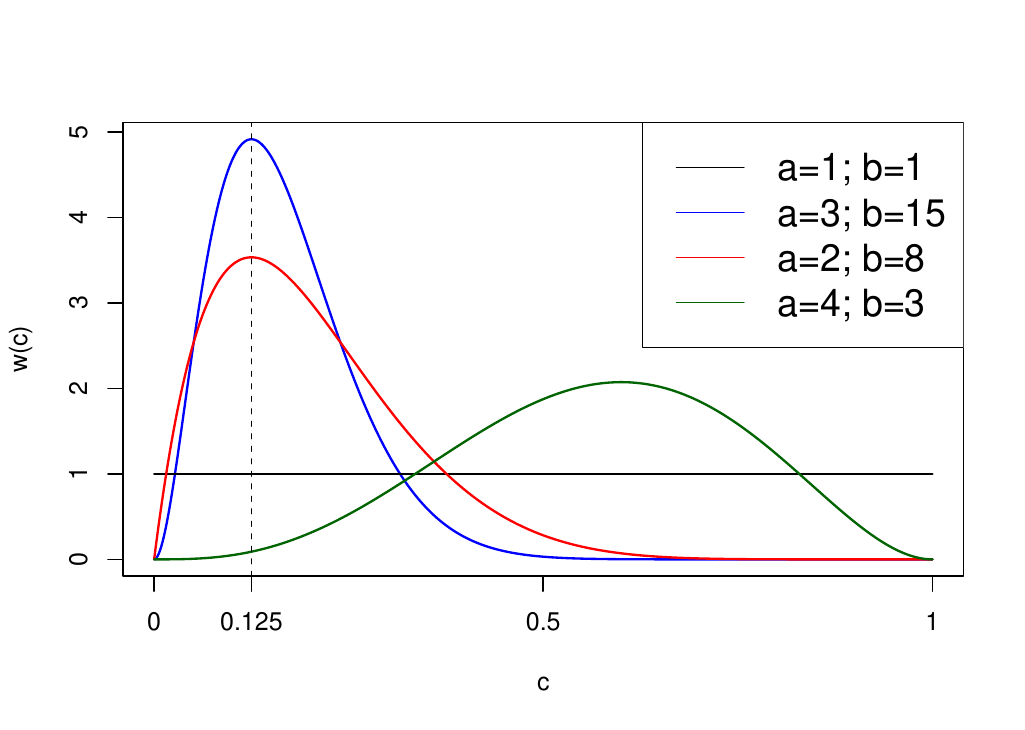}}
\caption{Beta distributions as the weight function\label{fig1}}
\end{figure}

There are some similarities between choosing a sensible weight function and the informative prior elicitation in Bayesian analysis: 1) a simple distribution is chosen to represent the best knowledge about the population, with parameters often chosen by vague heuristics; 2) some sensitivity analyses by varying the parameters are helpful to assess the robustness of results.  The difference is that, unlike Bayesian analysis, the distribution of the cutoff or cost ratio is not the goal for inference. The distribution relies on the information intrinsic to the clinical or scientific problem, but is usually external to the dataset at hand. \cite{tsalatsanis2010regret} proposed an elicitation process for a single optimal risk cutoff in the context of the decision curve analysis. We can adopt their process to formally elicit the distribution of risk cutoff, and it is briefly described as follows: 1) educate physicians with the risk-based decision framework. \cite{tsalatsanis2010regret} may help in this step; 2) present the physicians with patients from a real or hypothetical dataset with key covariates (e.g., age) and risk estimates that are representative of the population; 3) ask the physicians to make a treatment decision for each patient; 4) the decisions paired with the risk estimates can be used to construct the cutoff distribution.

Our proposal should not be confused with a different version of the weighted Brier score proposed in \cite{gerds2008performance}:
\begin{equation*}
    \frac{1}{n}\sum_{i=1}^{n} \{w_1 Y_i (1-\hat{r}_i)^2+ w_0(1-Y_i)\hat{r}^2_i\},
\end{equation*}
where $w_1$ and $w_0$ are chosen to assign different weights for cases and controls. This way of assigning weights is not coherent with the framework described in Section \ref{risk framework} and Appendix \ref{app: risk framework}; hence, it is harder to specify or interpret. In contrast, our proposed weighted Brier score stemmed from the integral representation and quantifies the clinical utility within the risk-based decision framework. 

Based on $n$ i.i.d. copies of validation data ($\hat{r}_i$, $Y_i$) from a cohort study, the weighted Brier score can be empirically evaluated: 
\begin{equation} \label{eq:bsw_emp}
    \begin{split}
       BS_w=&\frac{1}{n} \sum_{i=1}^{n} \ell_w(\hat{r}_i,Y_i) \\
        =&\frac{1}{n} \sum_{i=1}^{n}[(1-Y_i)m_w(\hat{r}_i)+Y_i[1-F_w(\hat{r}_i)+m_w(\hat{r}_i)-\mu_w]],
    \end{split}
\end{equation} where $F_w(\hat{r}_i)=\int_{0}^{\hat{r}_i}w(c)dc$, $m_w(\hat{r}_i)=\int_0^{\hat{r}_i}\{w(c)c\}dc$, and $\mu_w=\int_0^1\{w(c)c\}dc$. In Appendix \ref{app: bsw_emp}, we provide more details of equation (\ref{eq:bsw_emp}) and its implementation in the R package. 

\cite{pfeiffer2020estimating} considered a scenario in which the risk model is well-calibrated (i.e., $E[Y|\hat{r}(\bm{X})=\hat{r}(\bm{x})]=\hat{r}(\bm{x})$). Under this scenario, we have an alternative estimator, $BS^c_w:=\frac{1}{n} \sum_{i=1}^{n} \ell_w(\hat{r}_i,\hat{r}_i)$. The asymptotic distributions of $BS_w$ and $BS^c_w$ are derived in Appendix \ref{app: variance}, and show that when the model is well calibrated, $BS^c_w$ is more efficient than $BS_w$. This result is analogous to those in \cite{pfeiffer2020estimating} for the estimation of net benefit. Also see \cite{marsh2020statistical} for the estimation and inference of net benefit in a (matched) case-control study, which could be extended to weighted Brier score.

\subsection{Decomposition of the weighted Brier score and standardization}\label{sect decomp}

The Brier score assesses the overall prediction performance. Murphy's decomposition of the Brier score \citep{murphy1973new} explicitly demonstrates that the Brier score inherently measures calibration and discrimination. Based on the recent work on the generalization of Murphy's decomposition \citep{brocker2009reliability,siegert2017simplifying}, we decompose the weighted Brier score into calibration and discrimination components to better understand the impact of weighting. Before generalizing decomposition to the weighted Brier score, we first review the classic result of Murphy's decomposition.

Murphy's decomposition involves grouping in a similar way as the construction of the Hosmer–Lemeshow test \citep{hosmer1980goodness}, by forming groups based on deciles of $\hat{r}_i$. Typically, one divide the samples into $K$ groups based on ranked values of $\hat{r}_i$, and $\bar{Y}_k$ and $\bar{r}_k$ are the sample average of the observed event rate and the predicted risk within the $k^{th}$ group, respectively; $n_k$ is the total sample size of the $k^{th}$ group; $\hat{\pi}=\bar{Y}$ is the overall event rate in the sample. As shown by \cite{murphy1973new}, the Brier score can be decomposed as: 

\begin{equation*} 
         \begin{split}
        \frac{1}{n} \sum_{i}^{n} (\hat{r}_i-Y_i)^2 & \approx \underbrace{\sum_{k=1}^{K} \frac{n_k}{n} (\bar{Y}_k-\bar{r}_k)^2}_{MCB}-
        \underbrace{\sum_{k=1}^{K} \frac{n_k}{n} (\bar{Y}_k-\hat{\pi})^2}_{DSC}+\underbrace{\hat{\pi}(1-\hat{\pi})}_{UNC} \\
        &=MCB-DSC+UNC.
         \end{split}
\end{equation*}
The decomposition is exact when the predicted risk is discrete (e.g., $\hat{r}_i \in \{0,0.1,0.2,...,0.9,1\}$), and the grouping for the continuous predicted risk results in a minimal discrepancy hence the approximation is usually sufficient enough in practical applications. We adopt the terminology from \cite{dimitriadis2021stable} to name the three components: the miscalibration component ($MCB$), the discrimination component ($DSC$), and the uncertainty component ($UNC$). In some literature, the first two terms are also known as the reliability component ($REL$) and the resolution component ($RES$), respectively. 

$MCB$ quantifies the differences between the expected event rate ($\bar{r}_k$) and the observed event rate ($\bar{y}_k$). $\bar{y}_k$ is also known as the recalibrated risk in the $k^{th}$ group. $DSC$ measures the differences between the recalibrated risks and the overall event rate ($\hat{\pi}$) for the discrimination ability. A highly discriminative risk prediction model would have $\bar{y}_k$ largely different from $\hat{\pi}$, which leads to a higher $DSC$. If we make an uninformative prediction for each observation by using the overall event rate ($\hat{r}_i=\hat{\pi}$), $DSC$ would be 0 ($MCB$ would be 0 as well). $UNC$ measures the inherent difficulties in predicting a binary outcome. 

\cite{brocker2009reliability} generalized Murphy's decomposition of the Brier score to any strictly proper scoring rule. We apply the methodology to the weighted Brier score. In particular, we can decompose the expected value of the weighted Brier score as follows:

\begin{equation} \label{eq:decomp}
 E_{\bm{X},Y}[BS_w]
=\underbrace{E_{\bm{X}}[d\{\hat{r}(\bm{X}),\tilde{r}(\bm{X})\}]}_{MCB_w}-\underbrace{E_{\bm{X}}[d\{\pi,\tilde{r}(\bm{X})\}]}_{DSC_w}+\underbrace{\ell_w(\pi,\pi)}_{UNC_w}
\end{equation} 
where $\tilde{r}(\bm{X})=E_{Y|
\bm{X}}[Y|\hat{r}(\bm{X})]$ is the recalibrated risk; $d\{p,q\}=\ell_w(p,q)-\ell_w(q,q)$ is known as divergence; $\ell_w(p,q)=\int_{0}^{1} [cI_{[p\ge c]}(1-q)+(1-c)I_{[p< c]}q] w(c)dc$; $\pi=E[Y]$. We derive equation (\ref{eq:decomp}) in Appendix \ref{app: decomp}. One can verify that when the weight function, $w(c)$, is the PDF of uniform distribution over (0,1), we have $d(p,q)=\frac{1}{2}(p-q)^2$ and $l_w(p,p)=\frac{1}{2}(1-p)p$; thus, the classical Murphy's decomposition of the Brier score is a special case. 

The decomposition suggests that the weighted Brier score also measures discrimination and calibration; compared with the Brier score, by construction, each component additionally takes the clinical utility into account, which is numerically demonstrated in Section \ref{simulation}.

\subsection{Connection to the $H$ measure} \label{H}

This subsection connects the weighted Brier score with the $H$ measure \citep{hand2009measuring,hand2010evaluating} for two reasons. First, the $H$ measure's popularity in machine learning provides further support for our weighting method. Second, the $H$ measure's neglect of calibration, mirroring the broader field's over-emphasis on AUC at the expense of calibration in risk model evaluation \citep {van2019calibration}, underscores the value of the weighted Brier score, which considers both discrimination and calibration.

The $H$ measure quantifies discrimination ability for predicting the binary outcome in a validation dataset. The construction of the $H$ measure shares the same spirit as the risk-based decision framework discussed in Section \ref{risk framework} and Appendix \ref{app: risk framework} in terms of choosing a cutoff to balance different cost trade-offs. 

Instead of studying an estimated risk and a cutoff with a probability interpretation, \cite{hand2009measuring} considered a general real-valued score $Z \in \mathbb{R}$ with a cutoff $t$ in the prediction problem, and an overall cost $Q$ of binary decision as:

\begin{equation} \label{Q}
    \begin{split}
    Q(t;C_0,C_1) & := C_0P(Z>t,Y=0)+C_1P(Z<t,Y=1) \\
    & \propto c(1-\pi)(1-F_0(t))+(1-c)\pi F_1(t) := Q(t;c),
    \end{split}
\end{equation}
where $C_1$ and $C_0$ are costs of false negatives and false positives; because only the relative magnitude of the costs matters, one may replace $C_1$ and $C_0$ with $c:=C_0 / (C_1+C_0)$ hence the proportional sign appears in (\ref{Q}). $F_0$ and $F_1$ are the CDFs of $Z$ among controls and cases. Through differentiation, one can note that an optimal cutoff $t$ to minimize $Q(t;z)$ must satisfy the following equality:

\begin{equation} \label{G}
    G(t):=\frac{(1-\pi)f_1(t)}{(1-\pi)f_0(t)+\pi f_1(t)}=c,
\end{equation}
where $f_0$ and $f_1$ are the PDFs of $Z$ among controls and cases. That is, $G^{-1}(c)=\underset{t}{argmin}Q(t;c)$, assuming that function $G(.)$ is invertible. See \cite{hand2009measuring} for a relaxation in this invertible assumption. Note that $G(t)=P(Y=1|Z=t)$ is also known as the risk score, and the above discussion is in line with the optimality of the risk score \citep{mcintosh2002combining}.  

\cite{hand2009measuring} showed that the AUC can be rewritten as:

\begin{equation*}
    AUC:=\int_{0}^{1} TPR(t) d FPR (t)=1-\frac{1}{2\pi(1-\pi)}\int_{0}^{1} Q(t=G^{-1}(c);c)w^{AUC}(c) dc ,
\end{equation*}
where $
    w^{AUC}(c)=\{ (1-\pi) f_0(G^{-1}(c))+\pi f_1(G^{-1}(c))\}\lvert \frac{dG^{-1}(c)}{dc}\rvert$.

This implies that if we choose the cutoff optimally, the AUC is a function of a weighted average of the overall cost ($Q$), and the weight $w^{AUC}(c)$ is a function of the disease prevalence ($\pi$) and the PDFs of $Z$ among cases and controls. In essence, the derivation of $w^{AUC}(c)$ involves changing of variables from the $FPR(t)$ scale to the risk cutoff scale $c$. Recall that the cutoff is key in the risk-based decision framework from Section \ref{risk framework} and Appendix \ref{app: risk framework}. The challenge in interpreting $w^{AUC}(c)$ is consistent with the clinical irrelevance in summarizing the ROC by integrating $TPR(t)$ uniformly over $FPR(t)$. The clinical irrelevance of AUC motivates some notions of weighted AUC, which place weights on $FPR(t)$. The partial AUC is a special case of this type of weighted AUC \citep{pepe2003statistical,li2010weighted}. 

Instead of having the weight function depend on the distribution of $Z$ or placing weights on $FPR(t)$ scale, \cite{hand2010evaluating} suggested that one should place weights $w(c)$ on the risk cutoff scale according to the costs trade-offs specific to the classification problem of interest, and define the weighted averaged cost as $V:=\int_{0}^{1} Q(G^{-1}(c);c)w(c) dc$. The choice and interpretation of $w(c)$ coincide with the construction of the weighted Brier score in Section \ref{wBS def}. 

Furthermore, if the continuous score ($Z$) is a well-calibrated risk estimate, the weighted average cost ($V$) equals the expected value of the weighted Brier score. To show this equality, note that if $z=P(Y=1|Z=z)$ (i.e., well-calibrated), $G(.)$ is the identity function via Bayes' theorem hence $Q(t=c;c)=L(c)$ pointwise; integrating with the same weight, $w(c)$, we have $V=E[BS_w]$. 

\cite{hand2009measuring} defined the $H$ measure by a standardization step: $H=1-V/V_{max}$. When the distributions of $Z$ are equal in cases and controls (i.e., $F_1=F_0$), $V_{max}=(1-\pi)\int_{0}^{\pi}cw(c)dc+\pi\int_{\pi}^{1} (1-c)w(c)dc$, which is identical to the uncertainty component ($UNC_w$) from the decomposition of the weighted Brier score. Therefore, when $Z$ is a well-calibrated risk estimate, the $H$ measure equals the expected value of a weighted Brier score scaled by its uncertainty component. 

The similarities and differences of the $H$ measure and the scaled weighted Brier score are summarized as follows. The $H$ measure does not restrict the prediction score $Z$ to a risk estimate; when the score is a risk estimate, the weighted Brier score assesses the score with or without miscalibration, and in a special case of well-calibration, the expected value of the scaled weighted Brier score reduces to the $H$ measure. 
\section{Numerical examples} \label{simulation}
In this section, we use various numerical examples to illustrate the merits of the weighted Brier score for comparing different risk models in a validation dataset. For the first part, we present two quite artificial sets of examples (Set A and Set B) with an arbitrarily large sample size ($N=1,000,000$) to clearly demonstrate the impact of weighting on the expectation of weighted Brier score via its $DSC_w$ and $MCB_w$ components. The estimation of the weighted Brier score in a cohort study with a moderate sample size is straightforward, and the corresponding results are summarized in Appendix \ref{app:finite}. 
In the second part, we aim to demonstrate the weighted Brier score as a useful overall summary measure. We achieve this aim using a numerical example based on a real dataset with some simulated variables to mimic a potential dilemma in practice: comparing two risk prediction models when one model only outperforms the other model in discrimination or calibration but does not dominate both aspects simultaneously. Hereafter, we use $BS_w(a,b)$ to denote a weighted Brier score with the weight function from $Beta(a,b)$. In each example, multiple $BS_w(a,b)$ emphasized on a similar risk region are chosen as a sensitivity analysis.  
\subsection{Demonstration on the impact of weighting on $DSC_w$ and $MSC_w$}
\subsubsection{Set A: the impact on $DSC_w$}
In the first set, we generate three models with the same AUC value, but only one model is the best in this simulated scenario. 
The binary outcome ($Y$) is generated as a Bernoulli variable with $p=0.5$. Among cases ($Y=1$), $X \sim N(2,2)$ in Model 1, and $X \sim N(1,0.5)$ in Model 2. Among controls ($Y=0$), $X \sim N(0,1)$ for both Model 1 and Model 2. Via Bayes' theorem, we can calculate the predicted risk, $P(Y=1|X)$, denoted as $r_1$ and $r_2$, respectively, for Model 1 and Model 2. The predicted risk for Model 3, $r_3$, is generated based on $r_2$ as:
\begin{equation*}
        r_3= 
\begin{cases}
    expit(logit(r_2)+1),& \text{if } r_2 \geq 0.3\\
    expit(logit(r_2)-1),& \text{if } r_2 < 0.3
\end{cases}.
\end{equation*}

Construction of Model 3 with over-fitted risk is similar to the examples from \cite{hilden2014note} and \cite{pepe2015net}.

\begin{figure}[h]
\centerline{\includegraphics[scale=0.70]{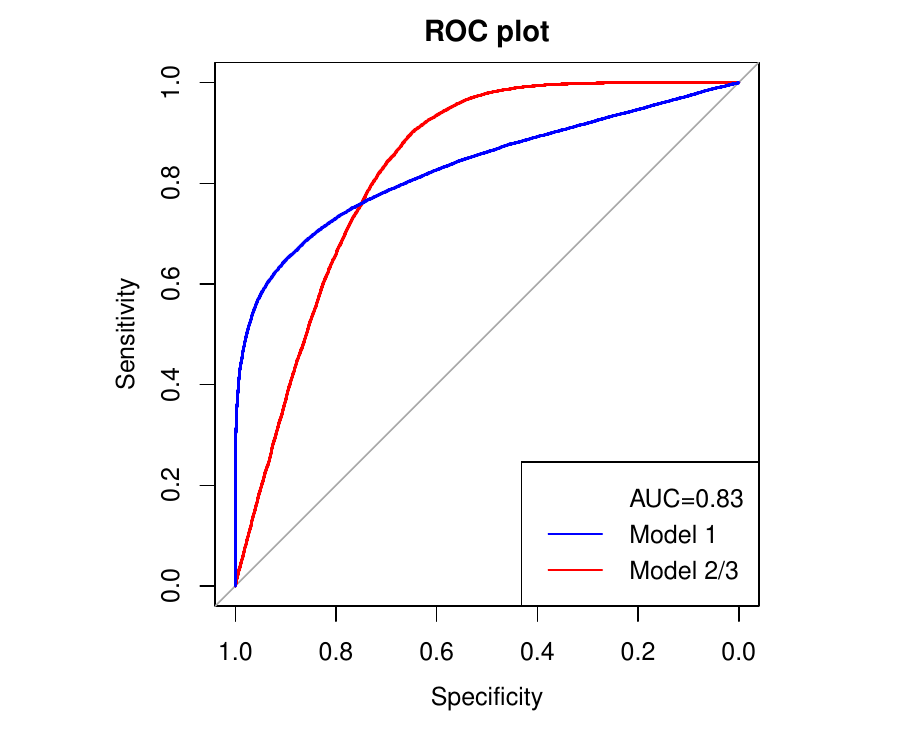}}
\caption{ROC plot of the three models\label{fig2}}
\end{figure}

\begin{center}
\begin{table}[h]
\caption{Comparing the three models with difference summary measures\label{tab2}}

\centering
\begin{tabular}{lrrr}
  \hline
 & Model 1 & Model 2 & Model 3 \\ 
  \hline
  AUC & 0.831 & 0.831 & 0.831 \\
$NB^{opt-in}$(0.3) & 0.327 & 0.384 & 0.384 \\
 IPA& 0.372& 0.372&0.288\\ 
  $BS_w$(1,1) & 0.078 & 0.078 & 0.089 \\ 
  \hspace{3mm}  $MCB_w$ & 0.000 & 0.000 & 0.010 \\ 
  \hspace{3mm}  $DSC_w$ & 0.046 & 0.046 & 0.046 \\ 
  $BS_w$(2,5) & 0.096 & 0.073 & 0.076 \\ 
  \hspace{3mm} $MCB_w$ & 0.000 & 0.000 & 0.003 \\ 
  \hspace{3mm} $DSC_w$ & 0.036 & 0.059 & 0.059 \\ 
  $BS_w$(4,8) & 0.110 & 0.084 & 0.087 \\ 
  \hspace{3mm} $MCB_w$ & 0.000 & 0.000 & 0.002 \\ 
  \hspace{3mm} $DSC_w$ & 0.049 & 0.074 & 0.074 \\ 
   \hline
\end{tabular}
\begin{tablenotes}%%[200pt]
\item $NB^{opt-in}$(0.3): the opt-in net benefit with cutoff at 0.3; $BS_w(a,b)$: a weighted Brier score with the weight function from $Beta(a,b)$; $MCB_w$ ($DSC_w$):  the miscalibration (discrimination) component of the weighted Brier score.  
\end{tablenotes}

\end{table}    

\end{center}

Based on the classic result of binormal ROC \citep{pepe2003statistical}, Model 1 and Model 2 have the same AUC, and due to the monotonic transformation, Model 3 and Model 2 have the identical ROC as shown in Figure \ref{fig2}. Similar to the AUC, the unweighted Brier score ($BS_w(1,1)$) or its scaled version, IPA \citep{kattan2018index}, cannot distinguish Model 1 and Model 2 as shown in Table \ref{tab2}. In clinical scenarios where a majority of the population chooses low-risk cutoffs (i.e., not treating a case is worse than treating a control), compared with Model 1, Model 2 has better clinical utility as demonstrated by the smaller values in the weighted Brier scores ($BS_w(2,5)$ or $BS_w(2,8)$) with additional weights placed on the lower cutoffs. The smaller weighted Brier score results from a higher $DSC_w$ in Model 2, which suggests better discrimination for these specific clinical scenarios. Loosely speaking, one may visualize that the additional weights are put on the high-sensitivity region on the top right of the ROC plot (more precisely, the low-risk region on the absolute risk scale). Because Model 1 and Model 2 are perfectly calibrated due to the correct model specification via the Bayes formula, any weight function would lead to $MCB_w=0$. 

The comparison between Model 2 and Model 3 aims to demonstrate the non-strictness of the net benefit measures (See Appendix \ref{app: risk framework} for the definitions of net benefit measures). Note that $NB^{opt-in}$(0.3) cannot distinguish an over-fitted/miscalibrated model (Model 3) from its perfectly calibrated counterpart (Model 2). On the other hand, as a strictly proper scoring rule, the weighted Brier scores pick Model 2 as the best model. 
\subsubsection{Set B: the impact on $MCB_w$}

In Set A, by comparing two models with the same AUC but crossed ROC, we demonstrate that the weighted Brier score can pick up the best model in terms of discrimination for the specific cost trade-offs, whereas the unweighted Brier score cannot. In Set B, we construct an example that the unweighted Brier score cannot distinguish two models with miscalibration in the lower or higher risk regions, but the weighted Brier score can meaningfully distinguish the miscalibration that matters the most for clinical decision-making. 

The binary outcome ($Y$) is again generated as a Bernoulli variable with $p=0.5$. For the True Model, among cases ($Y=1$), $X \sim N(1,1)$; among controls ($Y=0$), $X \sim N(0,1)$. Again, via Bayes' theorem, the predicted risk of the True model, $r_{T}$, can be calculated. Two miscalibrated models are constructed based on the True Model. In particular, an over-fitted in high-risk region (OH) model is constructed as:

\begin{equation*}
        r_{OH}= 
\begin{cases}
    expit(logit(r_T)+1),& \text{if } r_T \geq 0.5\\
    r_T,& \text{if } r_T < 0.5
\end{cases}.
\end{equation*}

An over-fitted in low-risk region (OL) model is constructed as: 
\begin{equation*}
        r_{OL}= 
\begin{cases}
    r_T,& \text{if } r_T \geq 0.5\\
    expit(logit(r_T)-1),& \text{if } r_T < 0.5
\end{cases}.
\end{equation*}

Again, we consider the scenarios that the risk cutoff for the majority of the population is smaller than 0.5. In these scenarios, we expect the miscalibration in the lower-risk region to have a more negative impact on the decision-making and, hence, the clinical utility of the model. Indeed, as shown in Table \ref{tab3}, the weighted Brier scores with more weights in the lower risk region ($BS_w(2,5)$ or $BS_w(4,8)$) are lower in the OH model than those in the OL model, indicating inferior clinical utility of the OL model. More specifically, the difference in $BS_w$ is due to the impact of weighting on $MCB_w$; as expected, the three models share the same $DSC_w$ regardless of the weights. Due to the symmetry in the data-generating mechanism, the unweighted Brier score ($BS_w(1,1)$) or its scaled version, IPA, shows the same level of miscalibration in terms of $MCB$.

\begin{center}
\begin{table}[h]
\caption{Comparing the three models with difference summary measures\label{tab3}}

\centering
\begin{tabular}{lrrr}
  \hline
 & True Model & OH Model & OL Model \\
\hline
 IPA& 0.204& 0.147&0.147\\ 

$BS_w$(1,1) & 0.099 & 0.107 & 0.107 \\ 
\hspace{3mm}  $MCB_w$ & 0.000 & 0.007 & 0.007 \\ 
\hspace{3mm}  $DSC_w$ & 0.025 & 0.025 & 0.025 \\ 
$BS_w$(2,5) & 0.107 & 0.108 & 0.122 \\ 
\hspace{3mm}  $MCB_w$ & 0.000 & 0.001 & 0.016 \\ 
\hspace{3mm}  $DSC_w$ & 0.026 & 0.026 & 0.026 \\ 
$BS_w$(4,8) & 0.124 & 0.124 & 0.141 \\ 
\hspace{3mm}  $MCB_w$  & 0.000 & 0.001 & 0.017 \\ 
\hspace{3mm}  $DSC_w$ & 0.035 & 0.035 & 0.035 \\ 
   \hline
\end{tabular}
\begin{tablenotes}%%[200pt]
\item  $BS_w(a,b)$: a weighted Brier score with the weight function from $Beta(a,b)$; $MCB_w$ ($DSC_w$):  the miscalibration (discrimination) component of the weighted Brier score.  
\end{tablenotes}

\end{table}    

\end{center}

\subsection{The weighted Brier score as a useful overall summary} 

Based on Prostate Cancer Active Surveillance Study (PASS) data, we demonstrate the weighted Brier score as a useful overall summary to incorporate clinical utility when evaluating rival models in a difficult scenario: some models are better in terms of discrimination, while the other models are better in terms of calibration. 
In PASS, risk models are built to predict high-grade cancer at each biopsy with a set of clinical covariates \citep{lin2017evaluating}. The clinical application of these risk models is to inform a risk-based decision on whether to opt out of a surveillance biopsy. A patient with a low risk of high-grade cancer for the next surveillance biopsy may skip the biopsy, which is invasive and may cause infection. However, skipping a surveillance biopsy may miss a progressive cancer. In general, missing a progressive cancer is several times more costly than an unnecessary invasive biopsy. 

Our goal is to compare rival risk models for the clinical application that relies on the framework described in Section \ref{risk framework} and Appendix \ref{app: risk framework}. Suppose the physicians agree that the cost of missing progressive cancer for a typical patient in this population ($C_1$) is about seven times more than the cost of an unnecessary invasive biopsy ($C_0$), which results in $c=1/8$. Thus, the net benefit, $NB^{opt-out}(1/8)$, is considered as a summary measure. To acknowledge the heterogeneity in the cost-benefit assessment, each patient in the population may consider a different risk threshold. Thus, we also consider several weight Brier scores (i.e., $BS_w(2,8)$ and $BS_w(3,15)$). Note that the modes of $Beta(2,8)$ and $Beta(3,15)$ are both 1/8, and the $95^{th}$ percentiles of $Beta(2,8)$ and $Beta(3,15)$ are approximately 0.43 and 0.33, respectively. That is, very few patients would consider a risk threshold higher than 0.4. Two weighted Brier scores are chosen here as a sensitivity analysis. Just for illustrative purposes, we also consider an irrational choice of weight, $Beta(4,3)$, which places the majority of its mass on the cutoffs greater than 0.5 (Figure \ref{fig1}), suggesting an unlikely patient's cost-benefit assessment that the cost of an unnecessary invasive biopsy is greater than the cost of missing progressive cancer. In addition, we also consider the AUC, the mean calibration \citep{van2019calibration}, the unweighted Brier score ($BS_w(1,1)$), and its scaled version, IPA \citep{kattan2018index}. 

The dataset consists of 3384 active surveillance biopsies from 1612 patients. We randomly split the patients into a training set and a validation set in a 1:1 ratio. In the training set, we fit a logistic regression model with high-grade cancer as the binary outcome (i.e., $Y=1$ for Gleason Group 2 or higher) and a set of clinical variables as the covariates, including prostate size, PSA, prior max cores ratio, number of prior negative biopsies, BMI, time since diagnosis and age. 

While the biopsies and their observed results ($Y$) in the data were obtained from the leading academic hospitals of the PASS sites, we consider a practical setting in which less accurate biopsy results were obtained from some hypothetical community clinics to highlight the potential dilemma when evaluating competing models. We simulate two sets of inaccurate results with different levels of accuracy for comparison purposes. In particular, we simulate $S1$ with $P(S1=0|Y=1)=0.25$ and $P(S1=1|Y=0)=0.05$; $S2$ with $P(S2=0|Y=1)=0.15$ and $P(S2=1|Y=0)=0.15$. The event rates of $Y$, $S1$, and $S2$ are 0.23, 0.2, and 0.31, respectively. The logistic regression with the same set of covariates is fitted with $S1$ or $S2$ as the outcomes. We call the logistic regressions with $Y$, $S1$, and $S2$ as Model Y, Model S1, and Model S2 respectively. 

Furthermore, we simulate an additional variable $X\sim N(0.7,1)$ in cases and $X \sim N(0,1)$ in controls. We fit another logistic regression model with $S2$ as the outcome and the same set of clinical variables plus the new variable $X$ as the covariates. We denote this model as Model S2+. We expect Model S2+ to have a higher AUC due to the additional predictor $X$ but less calibrated than Model Y due to the errors in $S2$.

\begin{center}

\begin{table}[h]

\caption{Comparing the four models for PASS data with different summary measures (95\% CI) \label{tab4}}
\small
\begin{tabular}{rllll}
  \hline
 & Model Y & Model S1 & Model S2 & Model S2+ \\ 
  \hline
AUC & 0.76 (0.73,0.79) & 0.76 (0.73,0.79) & 0.76 (0.73,0.79) & 0.78 (0.75,0.81) \\ 
  $O/E$ & 1.0 (0.95,1.1) & 1.1 (1,1.2) & 0.73 (0.67,0.8) & 0.73 (0.67,0.8) \\ 
  $BS_w(1,1)$ & 0.076 (0.07,0.082) & 0.077 (0.071,0.084) & 0.08 (0.076,0.085) & 0.078 (0.074,0.083) \\ 
  IPA & 0.16 (0.12,0.19) & 0.14 (0.11,0.16) & 0.11 (0.068,0.14) & 0.13 (0.087,0.16) \\ 
  $BS_w(2,8)$ & 0.088 (0.081,0.094) & 0.091 (0.085,0.096) & 0.10 (0.097,0.10) & 0.096 (0.092,0.10) \\ 
  $BS_w(3,15)$ & 0.084 (0.077,0.09) & 0.087 (0.081,0.092) & 0.099 (0.096,0.1) & 0.095 (0.091,0.098) \\ 
  $BS_w(4,3)$ & 0.089 (0.081,0.098) & 0.091 (0.082,0.10) & 0.091 (0.084,0.098) & 0.09 (0.083,0.097) \\ 
  $NB^{opt-out}(1/8)$ & 0.16 (0.099,0.23) & 0.10 (0.047,0.15) & 0.013 (0.0010,0.024) & 0.037 (0.023,0.049) \\ 
   \hline
\end{tabular}

\begin{tablenotes}%%[200pt]
\item $BS_w(a,b)$: a weighted Brier score with the weight function from $Beta(a,b)$; $O/E$: observed event rate over expected average risk, i.e., the mean calibration \citep{van2019calibration}
\end{tablenotes}

\end{table}    

\end{center}

\begin{figure}[h]
\centerline{\includegraphics[scale=0.750]{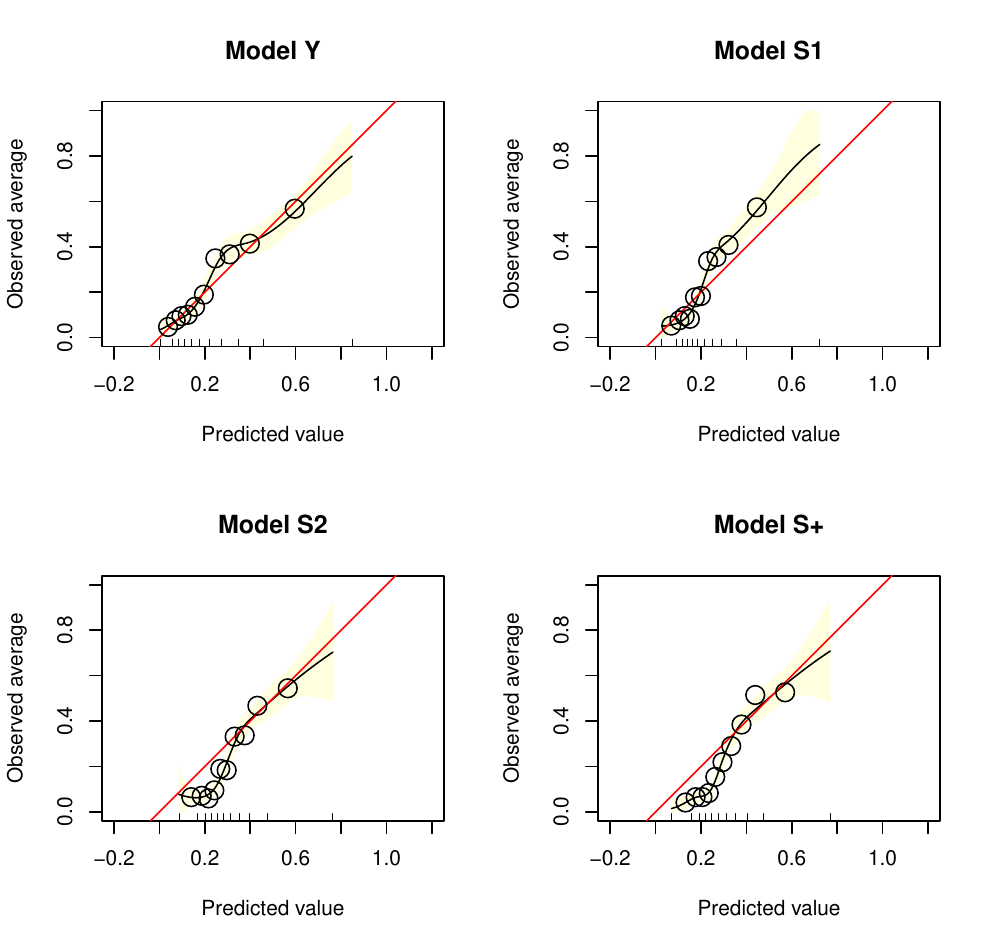}}
\caption{Calibration plot of the four models\label{fig3}}
\end{figure}

The predicted risks of high-grade cancer from each model are evaluated in the validation set with $Y$ as the outcome. The summary measures are presented in Table \ref{tab4}. We construct the confidence intervals using the bootstrap approach with each individual patient as the resampling unit to account for the correlation. Model Y, Model S1, and Model S2 have almost identical AUC because the non-differential errors in the binary outcome would not change the rank of predicted risks \citep{neuhaus1999bias}. Due to the errors in the outcome, Model S1, Model S2, and Model S2+ are not well calibrated, the mean calibration, $O/E$, in Table \ref{tab4}. As shown in Figure \ref{fig3}, Model S1 is less calibrated and tends to underestimate the risk in the high-risk region, and Model S2 and Model S2+ are less calibrated and tend to overestimate the risk in the low-risk region. 

\begin{figure}[h]
\centerline{\includegraphics[scale=0.6]{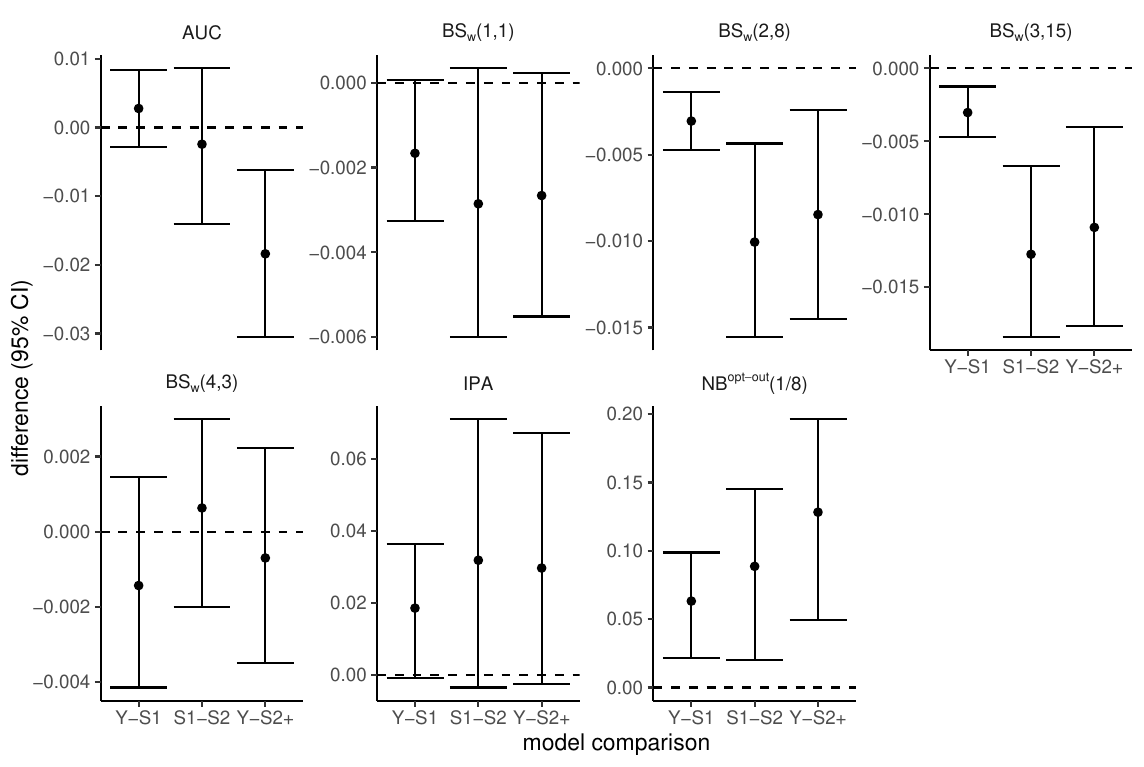}}
\caption{Selected pairwise comparisons between Model Y and Model S1 (Y-S1), Model S1 and Model S2 (S1-S2), Model Y and Model S2+ (Y-S2+). Differences (y-axis) in various summary measures are presented in each panel for each pairwise comparison (x-axis). The difference in each summary measure is the summary measure of the former model minuses the summary of the latter model. The positive difference in the AUC, IPA, and $NB^{opt-out}(1/8)$ suggests the former model is better. The negative difference in $BS_w(.,.)$ suggests the former model is better overall.\label{fig4}}
\end{figure}

We visualize some selected pairwise comparisons among four models in Figure \ref{fig4}. Overall, $NB^{opt-out}(1/8)$, $BS_w(2,8)$ and $BS_w(3,15)$ clearly indicate that Model Y is better than Model S1 or Model S2+; Model S1 is better than Model S2. Although $BS_w(1,1)$ and IPA generally indicate similar conclusions, their confidence intervals cover 0. 

Specifically, with more weight placed on the low-risk region (e.g., with $BS_w(2,8)$ instead of $BS_w(1,1))$, the increasingly lower value in $BS_w$ in Model S1 compared with Model S2 suggests stronger evidence for the superiority of Model S1 over Model S2. This is expected because the heavier weight in the low-risk region amplifies the poorer calibration of Model S2 in the region, which significantly impacts the decision-making with the cost trade-off consideration. On the other hand, the positive difference in $BS_w(4,3)$ reflects a better calibration of Model S2 on the higher-risk region although the difference is much more subtle due to the small portion of patients in the high-risk region. As noted earlier, this comparison based on $BS_w(4,3)$ is irrational and not clinically meaningful when most of the cutoffs lay in the lower-risk region.  

\begin{figure}[h]
\centerline{\includegraphics[scale=0.7]{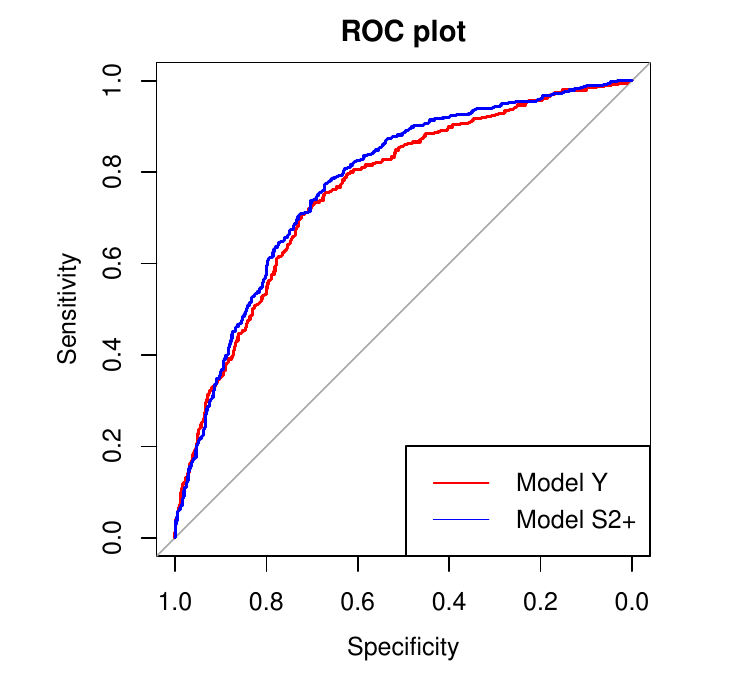}}
\caption{Model S2+ appears better in terms of discrimination.\label{fig ROC}}
\end{figure}

Compared with Model Y, Model S2+ appears to be better in discrimination in terms of the ROC (see Figure \ref{fig ROC} and the first panel of Figure \ref{fig4}) but less calibrated (see Figure \ref{fig3} and $O/E$ in Table \ref{tab3}) in the validation set, so we are in a dilemma of choosing a better model between these two models. The lower weighted Brier scores suggest that overall, Model Y is a better model in the context of the intended clinical application of the risk prediction model and cost trade-off in the patient population. This comparison highlights the importance of calibration for risk-based decisions, in particular in the relevant risk region, which may overshadow a good discriminatory ability.

\section{Discussion}
In this paper, we present a new clinically relevant summary measure for risk prediction models, namely, the weighted Brier score. The classic Brier score is recognized in the field as a useful overall summary measure for both discrimination and calibration but faces criticism for its inability to measure the clinical utility \citep{assel2017brier}. We analytically show this caveat of the classic Brier score and further construct the weighted Brier score that can meaningfully measure the clinical utility. The weighted Brier score shares the same decision-theoretic framework as the popular summary measure of net benefit. Specifying a distribution on the cutoffs for $BS_w$ is more natural for the population-level summary measure than requiring a single cutoff for the whole population, as the population of interest is often heterogeneous in its cost-benefit profile. 

The decision-theoretic framework involves a priori and vague knowledge about a particular application external to the data at hand. When equipped with rationally selected weights and coupled with sensitivity analyses as demonstrated in the numerical examples, the weighted Brier score offers an initial assessment and valuable information on the relative clinical utility of multiple risk prediction models. This knowledge facilitates the selection of a handful of promising models for further evaluation, motivating subsequent studies with longitudinal follow-up to objectively gather data regarding cost-effectiveness.

Our work utilizes many results from the study of proper scoring rules, which is most familiar to the weather forecast community. We additionally connect our work to the $H$ measure proposed by \cite{hand2009measuring}, which has been widely cited and utilized across various applications in the machine learning field, such as benchmarking credit scoring systems in finance \citep{lessmann2015benchmarking}. We hope this work also stimulates future methodology development in the field of medical risk prediction. To demonstrate this point on methodology development, we provide a quick example of constructing a weighted version of Spiegelhalter's Z-statistic \citep{spiegelhalter1986probabilistic,rufibach2010use} in Appendix \ref{app:wSz}. 

One interesting and important future direction is on the evaluation of prediction models for time-to-event outcomes, particularly in the context of flexible machine learning algorithms with their output being individualized survival curves \citep[e.g.,][]{haider2020effective,pmlr-v151-rindt22a,qi2023effective}. Along with adequate handling of censoring, clinically meaningful characterization of calibration, discrimination, and clinical utility for the prediction of time-to-event outcome warrants follow-up research in the lens of proper scoring rules. These predictions are probabilistic forecasting \citep{gneiting2014probabilistic}, and there has been active research in this area \citep{ehm2016quantiles,gneiting2023model}. Such follow-up research may generalize the metrics related to commonly studied t-year risks and the concordance probability \citep{heagerty2005survival} or its recent variants \citep{blanche2019c,devlin2020measuring,devlin2021concordance}.

\section*{Acknowledgments}
The work is supported by grants R01 CA236558 and U24 CA086368 awarded by the National Institutes of Health. The authors also thank the anonymous referee and the Associate Editor for their comments.

\bibliography{mybib}
\section*{Appendix}
\appendix
\section{More details on clinical utility in risk-based framework} \label{app: risk framework} 
We provide additional details on the risk-based framework and explicitly link the net benefit measures with $L(c):=cP(r>c,Y=0)+(1-c)P(r\leq c,Y=1)$. Recall that we recommend treating a patient if $r$ is greater than a cutoff $c$ (i.e., $I[r>c]$, where $I[.]$ is the indicator function and 1 means recommending a treatment).
 
The costs of 4 combinations of treatment recommendations and disease status are denoted as $C_{TP}$, $C_{FN}$, $C_{FP}$, and $C_{TN}$. The overall cost of this risk-based treatment recommendation is
\begin{equation} \label{overall cost}
    C_{TP}P(r>c,Y=1)+C_{FN}P(r\leq c,Y=1)+C_{FP}P(r>c,Y=0)+C_{TN}P(r\leq c,Y=0) .
\end{equation}

These costs, $C_{TP}$, $C_{FN}$, $C_{FP}$ and $C_{TN}$, are parts of the cost-benefit analysis of the clinical problem such as taking statin to prevent cardiovascular disease. Although these costs are not parameters of a risk model itself, evaluating the clinical utility of a risk model takes these costs into account. To simplify, we can rewrite the overall cost in terms of relative benefits or costs by setting a reference treatment policy. When the default treatment policy is to treat no one (treat-none), the overall cost is 
\begin{equation} \label{eq cost treat-none} 
C_{FP}P(Y=1)+C_{TN}P(Y=0).  
\end{equation}
Subtract (\ref{overall cost}) from (\ref{eq cost treat-none}), we get the relative benefit of the risk-based opt-in treatment policy is
\begin{equation} \label{nb-in}
    B\times P(r>c,Y=1)-C\times P(r>c,Y=0),
\end{equation}
where $B=C_{FN}-C_{TP}$ is the relative benefit gained for treating a case; $C=C_{FP}-C_{TN}$ is the relative cost incurred for treating a control.

Similarly, when the default treatment policy is to treat everyone (treat-all), the relative benefit of the risk-based opt-out treatment policy:
\begin{equation} \label{nb-out}
    -B\times P(r\leq c,Y=1)+C\times P(r\leq c,Y=0),
\end{equation}
where $B=C_{FN}-C_{TP}$ is also the relative benefit lost for not treating a case; $C=C_{FP}-C_{TN}$ is also the relative cost avoided for not treating a control.

Lastly, when the reference policy is ideal: treating all the cases and not treating any control, the relative cost of the risk-based treatment policy is 
\begin{equation} \label{nc} 
    C_1P(r\leq c,Y=1)+C_0P(r>c,Y=0),
\end{equation}
where $C_1=C_{FN}-C_{TP}$ is the relative cost incurred for not treating a case; $C_0=C_{FP}-C_{TN}$ is the relative cost incurred for treating a control.

Researchers from various fields \citep[e.g.,][]{pauker1975therapeutic,elkan2001foundations} noted that the optimal risk cutoff to maximize the relative benefits (\ref{nb-in}) and (\ref{nb-out}), or to minimize the relative cost (\ref{nc}) is 

\begin{equation} \label{opt-c}
c=\frac{C_{FP}-C_{TN}}{C_{FP}-C_{TN}+C_{FN}-C_{TP}}=\frac{C}{C+B}=\frac{C_0}{C_0+C_1}.
\end{equation}

This important result shows that there is a one-to-one relationship between the optimal risk cutoff and the cost ratio. In the later sections, we use this result to interpret the Brier score or its weighted versions.  

Finally, by combining (\ref{opt-c}) and (\ref{nb-in}) and with further normalization by dividing $B$, we arrive at the opt-in net benefit:
\begin{equation} \label{NB-in}
\begin{split}
    NB^{opt-in}(c):&=P(r>c,Y=1)-\frac{c}{1-c}P(r>c,Y=0) \\
    &=TPR(c)\pi-\frac{c}{1-c}FPR(c)(1-\pi) \ ,
\end{split}
\end{equation}
where $TPR(c)=P(r>c|Y=1)$ is the true positive rate at the cutoff $c$, $FPR(c)=P(r>c|Y=0)$ is the false positive rate at the cutoff $c$, and $\pi=P(Y=1)$. The above derivation from the overall costs in (\ref{overall cost}) to the $NB^{opt-in}$ in (\ref{NB-in}) suggests that if we choose the treat-none policy as the reference and the risk cutoff optimally to minimize the total costs, we can evaluate a risk prediction model in term of $TPR(c)$, $FPR(c)$ and $\pi$ at the risk cutoff $c$ corresponding to a cost ratio. Thus, the net benefit quantifies the clinical utility of risk models in the risk-based decision framework that incorporates the costs. 

Similarly, with normalization terms, $C$ or $\frac{1}{C_1+C_0}$, we formulate the opt-out net benefit ($NB^{opt-out}(c)$) or the cost-weighted misclassification error ($L(c)$) when the reference policy is treat-all or the ideal policy respectively: 
\begin{equation} \label{NB-out}
\begin{split}
     NB^{opt-out}(c):&=P(r\leq c,Y=0)-\frac{1-c}{c}P(r\leq c,Y=1) \\
     &=TNR(c)(1-\pi)-\frac{1-c}{c}FNR(c)\pi
\end{split}
\end{equation}

or 
\begin{equation} \label{CWM}
\begin{split}
    L(c)&:=cP(r>c,Y=0)+(1-c)P(r\leq c,Y=1) \\
    &= cFPR(c)(1-\pi)+(1-c)FNR(c)\pi \ ,
\end{split}
\end{equation}
where $TNR(c)=P(r\leq c|Y=0)$ is the true negative rate at the cutoff $c$, and $FNR(c)=P(r>c|Y=0)$ is the false positive rate at the cutoff $c$.

The above constructions and connections are summarized in Table \ref{tab1}, which is adopted and extended from Table 6.5 of \cite{pfeiffer2017absolute}. 

\begin{center}
\begin{table*}[h]%
\caption{Cost matrix of a decision problem with two health states and two interventions\label{tab1}}
\centering
\begin{tabular*}{\textwidth}{cccccc}
\toprule
& & &  \multicolumn{3}{@{}c@{}@{}}{\textbf{Costs/benefits relative to a reference policy }} \\\cmidrule{4-6}
\textbf{Treatment} & \textbf{Disease State} &  \textbf{Costs} & \multicolumn{1}{@{}c@{}}{\textbf{Treat-none}}  & \textbf{Treat-all}  & \textbf{Ideal policy}  \\
$I_{[r>c]}$& $Y$ & \\
\midrule
1 & 1  & $C_{TP}$   & $B=C_{FN}-C_{TP}$ & 0 &  0\\
0 & 1  & $C_{FN}$   & 0 & $B=C_{FN}-C_{TP}$  & $C_1=C_{FN}-C_{TP}$\\
1 & 0  & $C_{FP}$   & $C=C_{FP}-C_{TN}$ & 0  & $C_0=C_{FP}-C_{TN}$\\
0 & 0  & $C_{TN}$   & 0  & $C=C_{FP}-C_{TN}$  &  0\\
\midrule
& & & $NB^{opt-in}(c)$ & $NB^{opt-out}(c)$ & $L(c)$\\
\bottomrule
\end{tabular*}
\begin{tablenotes}%%[341pt]
\item Abbreviation: $C_{TP}$, $C_{FN}$, $C_{FP}$ and $C_{TN}$, costs of true positive, false negative, false positive and true negative respectively; $B$, relative benefit gained for treating a case or relative benefit lost for not treating a case; $C$, relative cost incurred for treating a control or relative cost avoided for not treating a control; $C_1$, relative cost incurred for not treating a case; $C_0$, relative cost of incurred for treating a control; $NB^{opt-in}(c)$ ($NB^{opt-out}(c)$), the net benefit of a
risk-based opt-in (opt-out) treatment policy; $L(c)$, cost-weighted misclassification error.
\end{tablenotes}
\end{table*}
\end{center}

When comparing competing models (e.g., Model A $\hat{r}_A$ and Model B $\hat{r}_B$), all three related measures give the same conclusion; that is, $I[L_A(c)<L_B(c)]=I[NB_{A}^{opt-in}(c)>NB_{B}^{opt-in}(c)]=I[NB_{A}^{opt-out}(c)>NB_{B}^{opt-out}(c)]$. It is easy to verify that the following relationship between $L(c)$, $NB^{opt-in}(c)$ and $NB^{opt-out}(c)$:
\begin{equation} \label{equality}
    L(c)=(1-c)[\pi-NB^{opt-in}(c)]=c[1-\pi-NB^{opt-out}(c)] \ .
\end{equation}

Therefore, along with the net benefit measures, $L(c)$ is an assessment measure that quantifies the clinical utility of risk prediction models at a specific $c$ corresponding to a cost ratio. In practice, specifying the optimal cutoff $c$ might be challenging. The decision curve, plotting the net benefit across $c$, can be considered a sensitivity analysis to examine the clinical utility at different $c$ \citep{vickers2006decision}. \cite{kerr2016assessing} emphasized that the net benefit is a summary of population-level model performance and assumes a constant $c$ in the population, which may be oversimplified. For example, in a population of cancer patients facing a decision to receive an aggressive surgery treatment, relatively young patients may have a lower risk cutoff of cancer progression to receive surgery than old patients due to different cost ratios (e.g., 0.1 in the young and 0.2 in the old), since the survival benefit of successfully treating a young patient is greater than an older patient in general. That is, a young patient with a predicted risk of 0.15 may be recommended for surgery, and an old patient with the same predicted risk of 0.15 may be recommended for a conservative treatment. We can compare the net benefits or $L(c)$ separately at 0.1 or 0.2 for two competing models predicting cancer progression; however, the corresponding net benefits or $L(c)$ assume everyone in the population has the same risk cutoff. Another example is a risk-based decision for invasive biopsy in a patient population. Individuals in the target population may have a range of susceptibility to infection due to the biopsy, which implies a distribution of cutoffs.

\section{$l_w(\hat{r}_i,Y_i)$}\label{app: bsw_emp}

In this section, we provide the details on calculating $l_w(\hat{r}_i,Y_i)$ empirically. 

\begin{equation*}
    \begin{split}
    l_w(\hat{r}_i,Y_i)&=\int_{0}^{1} \{cI[\hat{r}_i>c, Y_i=0]+(1-c)I[\hat{r}_i\leq c, Y_i=1] w(c)\}dc \\
    &=(1-Y_i)\int_{0}^{\hat{r}_i}c w(c)dc+Y_i\int_{\hat{r}_i}^{1}(1-c)w(c)dc \\
    &=(1-Y_i)m_w(\hat{r}_i) + Y_i\{1-\int_{0}^{\hat{r}_i}w(c)dc+\int_{0}^{\hat{r_i}}cw(c)dc-\int_{0}^{1}cw(c)dc\} \\
    &=(1-Y_i)m_w(\hat{r}_i)+Y_i(1-F_w(\hat{r}_i)+m_w(\hat{r}_i)+\mu_w), 
    \end{split}
\end{equation*}
where $F_w(\hat{r}_i)=\int_{0}^{\hat{r}_i}w(c)dc$, $m_w(\hat{r}_i)=\int_0^{\hat{r}_i}\{w(c)c\}dc$, and $\mu_w=\int_0^1\{w(c)c\}dc$.

When $w(c)$ is the PDF of a Beta distribution of $Beta(a,b)$, we have $\mu_{w}=\frac{a}{a+b}$ and 
\begin{equation*}
    \begin{split}
    m_w(\hat{r}_i):=&\int_{0}^{\hat{r}_i}c w(c)dc\\
    =&\int_{0}^{\hat{r}_i}c \frac{c^{a-1}(1-c)^{b-1}}{B(a,b)} dc \\
    =&\frac{B(a+1,b)}{B(a,b)}\int_{0}^{\hat{r}_i} \frac{c^{a}(1-c)^{b-1}}{B(a+1,b)} dc \\
    =&\frac{B(a+1,b)}{B(a,b)} F(a+1,b;\hat{r}_i),
    \end{split}
\end{equation*}
where $B(a,b)$ is the Beta function which can be obtained from \texttt{beta} function from the R package; $F(a,b; \hat{r}_i)$, the CDF of Beta distribution, can be obtain from \texttt{pbeta} function from the R package. Specifically, the following R function can be used to evaluate the weighted Brier score: 

\begin{verbatim}
#' A function for the weighted Brier score
#' 
#' This function implements the weighted Brier score with a Beta distribution as 
#' the weighting function. 
#' @param y a vector of binary outcomes
#' @param r the corresponding vector of risk estimates
#' @param a,b  the parameters for the Beta distribution/function
#' 
BS_w <- function (y, r, a, b) {
  l_w <- ((1 - y) * pbeta(r, shape1 = a + 1, shape2 = b) * beta(a = a + 1, b = b) +
            y * (1 - pbeta(r, shape1 = a, shape2 = b + 1)) * beta(a = a, b = b + 1)
          ) / beta(a = a, b = b)  
  return(mean(l_w))
}
\end{verbatim}

\section{Asymptotic distributions of $BS_w$ and $BS^c_w$} \label{app: variance}

In this section, we use the Central Limited Theorem (CLT) to show the asymptotic normality of the $BS_w=\frac{1}{n}\sum_{i=1}^n\ell_w(\hat{r}_i,Y_i)$ and $BS^c_w=\frac{1}{n}\sum_{i=1}^n\ell_w(\hat{r}_i,\hat{r}_i)$. We calculate and compare their variances with and without the well-calibrated assumption: $E_{Y_i|\bm{X}}[Y_i|\hat{r}(\bm{X_i})=\hat{r}(\bm{x_i})]=\hat{r}(\bm{x_i})$. 

To evoke the CLT, we assume that

\begin{enumerate}
    \item The outcome $Y_i$ is binary; the risk model $\hat{r}_i:=\hat{r}(\bm{X_i})$, a fixed function of predictors $\bm{X_i}$, is bounded by 0 and 1. 
    \item $\{(\hat{r}_i,Y_i)\}_{i=1,...,n}$ are i.i.d.
    \item The weight function $w(c)$ is a PDF with a finite range.  
\end{enumerate}

Under these three conditions, we note that $\{\ell_w(\hat{r}_i,Y_i)\}_{i=1,...,n}$ and $\{\ell_w(\hat{r}_i,\hat{r}_i)\}_{i=1,...,n}$ are both i.i.d.; they also have finite means and variances because $\ell_w(.,.)$ is a function with finite range and two arguments being bounded random variables.

\subsection{$BS_w=\frac{1}{n} \sum_{i=1}^{n} \ell_w(\hat{r}_i,Y_i)$} 

Denote $A(\hat{r}_i):=\int_{\hat{r}_i}^{1}(1-c)w(c)dc=1-F_w(\hat{r}_i)+m_w(\hat{r}_i)-\mu_w$ and $B(\hat{r}_i):=\int_{0}^{\hat{r}_i} cw(c)dc=m_w(\hat{r}_i)$.

From the CLT, we have
\begin{equation*}
    \sqrt{n}(BS_w-E[\ell_w(\hat{r}_i,Y_i)]) \xrightarrow[]{d} N(0,\sigma^2),
\end{equation*}
where 
\begin{equation*}
\begin{split}
    \sigma^2 &=var(\ell_w(\hat{r}_i,Y_i))\\
    &=E[\ell_w(\hat{r}_i,Y_i)^2]-E[\ell_w(\hat{r}_i,Y_i)]^2 \\
    &=E[\{Y_i A(\hat{r}_i)+(1-Y_i)B(\hat{r}_i)\}^2]-E[BS_w]^2 \\
    &=E[Y_i A(\hat{r}_i)^2+(1-Y_i)B(\hat{r}_i)^2]-E[BS_w]^2 .
\end{split}
\end{equation*}
Note that we used $Y_i=Y^2_i$, $(1-Y_i)=(1-Y_i)^2$ and $Y_i(1-Y_i)=0$. 

Under the well-calibration assumption, we have a new expression for the variance denoted as $\sigma^2_0$: 
\begin{equation*}
    \sigma^2_0=E[\hat{r}_i A(\hat{r}_i)^2+(1-\hat{r}_i)B(\hat{r}_i)^2]-E[BS_w]^2.
\end{equation*}

Conditional on $\hat{r}_i$, the empirical estimator of $\sigma^2_0$:
\begin{equation*}
    \begin{split}
    \sigma^2_{0n}&=1/n \sum_{i=1}^{n}[\hat{r}_i A(\hat{r}_i)^2+(1-\hat{r}_i)B(\hat{r}_i)^2-\{\hat{r}_iA(\hat{r}_i)+(1-\hat{r}_i)B(\hat{r}_i)\}^2]\\
    &=1/n \sum_{i=1}^{n}\hat{r}_i(1-\hat{r}_i)[A(\hat{r}_i)-B(\hat{r}_i)]^2,  
    \end{split}
\end{equation*} which is used in the weighted Spiegelhalter’s Z-statistic in (\ref{wSz}). 

\subsection{$BS^c_w=\frac{1}{n} \sum_{i=1}^{n} \ell_w(\hat{r}_i,\hat{r}_i)$} 

Under the well-calibration assumption, we have 
\begin{equation*}
    \begin{split}
        E[\ell_w(\hat{r}_i,Y_i)]=&E_{\bm X_i}[E_{Y_i|\bm X_i}\{\ell_w(\hat{r}(\bm X_i),Y_i)|\bm X_i\}] \\
        =&E_{\bm X_i}[\ell_w(\hat{r}(\bm X_i),E_{Y_i|\bm X_i}\{Y_i|\bm X_i\})] \\
        =&E_{\bm X_i}[\ell_w(\hat{r}(\bm X_i),\hat{r}(\bm X_i))]. 
    \end{split}
\end{equation*}

Note that the second equation holds because $\ell_w(\hat{r}(\bm X_i), Y_i)$ is linear in $Y_i$; the third equation is a result of the well-calibration assumption. Thus, $E[\ell_w(\hat{r}_i,\hat{r}_i)]= E[\ell_w(\hat{r}_i,Y_i)]=E[BS_w]$.

From the CLT, we have
\begin{equation*}
    \sqrt{n}(BS^c_w-E[\ell_w(\hat{r}_i,Y_i)]) \xrightarrow[]{d} N(0,\sigma^2_c),
\end{equation*}
where 
\begin{equation*}
\begin{split}
    \sigma^2_c &=var(\ell_w(\hat{r}_i,\hat{r}_i))\\
    &=E[\ell_w(\hat{r}_i,\hat{r}_i)^2]-E[\ell_w(\hat{r}_i,\hat{r}_i)]^2 \\
    &=E[\{\hat{r}_i A(\hat{r}_i)+(1-\hat{r}_i)B(\hat{r}_i)\}^2]-E[BS_w]^2 .
\end{split}
\end{equation*}

Make a comparison between the variances of $BS_w$ and $BS^c_w$ under the well-calibration assumption:

\begin{equation*}
    \sigma^2_0-\sigma^2_c=E[\hat{r}_i(1-\hat{r}_i)\{A(\hat{r}_i)-B(\hat{r}_i)\}^2].
\end{equation*}

Note that $A(\hat{r}_i)-B(\hat{r}_i)=1-F_w(\hat{r}_i)-\mu_w \neq 0$ for some $\hat{r}_i$; thus $\sigma^2_0-\sigma^2_c>0$ for non-generative cases (i.e. $\hat{r}_i$ is not constant 0 or 1). 

Therefore, under the well-calibration assumption, $BS^c_w$ is more efficient than $BS_w$.

\section{Decomposition of the expected value of weighted Brier score} \label{app: decomp}

We want to show that 
\begin{equation*} 
 E[BS_w]=\underbrace{E_{\bm{X}}[d(\hat{r}(\bm{X}),\tilde{r}(\bm{X}))]}_{MCB_w}-\underbrace{E_{\bm{X}}[d(\pi,\tilde{r}(\bm{X}))]}_{DSC_w}+\underbrace{\ell_w(\pi,\pi)}_{UNC_w}, 
\end{equation*}
where $\tilde{r}(\bm{X})=E_{Y|
\bm{X}}[Y|\hat{r}(\bm{X})]$ is the recalibrated risk; $d(p,q)=\ell_w(p,q)-\ell_w(q,q)$ is known as divergence; $\ell_w(p,q)=\int_{0}^{1} [cI_{[p\ge c]}(1-q)+(1-c)I_{[p< c]}q] w(c)dc$; $\pi=E[Y=1]$.

The decomposition is a telescoping sum that adds and subtracts $\ell_w(\pi,\pi)$ (a fixed function of $\pi$) and $E_{\bm{X}}[E_{Y|\bm{X}}\{\ell_w(\tilde{r}(\bm{X}),Y)\}]$:
\begin{equation*}
\begin{split}
 E[wBS]=&E[\ell_w(\hat{r},Y)]\\
 =&E[\ell_w(\hat{r},Y)-\ell_w(\tilde{r},Y)] - E[\ell_w(\pi,\pi)- \ell_w(\tilde{r},Y)]+\ell_w(\pi,\pi) \\
 =&E_{\bm{X}}[E_{Y|\bm{X}}\{\ell_w(\hat{r}(\bm{X}),Y)-\ell_w(\tilde{r}(\bm{X}),Y)|\bm{X}\}] - E_{\bm{X}}[E_{Y|\bm{X}}\{\ell_w(\pi,\pi)- \ell_w(\tilde{r}(\bm{X}),Y)|\bm{X}\}]+\ell_w(\pi,\pi) \\
 =&E_{\bm{X}}[\ell_w(\hat{r}(\bm{X}),\tilde{r}(\bm{X}))-\ell_w(\tilde{r}(\bm{X}),\tilde{r}(\bm{X}))] - E_{\bm{X}}[\ell_w(\pi,\tilde{r}(\bm{X}))-\ell_w(\tilde{r}(\bm{X}), \tilde{r}(\bm{X}))]+\ell_w(\pi,\pi) \\
=&\underbrace{E_{\bm{X}}[d(\hat{r}(\bm{X}),\tilde{r}(\bm{X}))]}_{MCB_w}-\underbrace{E_{\bm{X}}[d(\pi,\tilde{r}(\bm{X}))]}_{DSC_w}+\underbrace{\ell_w(\pi,\pi)}_{UNC_w}.
\end{split}
\end{equation*} 
Note that in the 4th line, we used the following identities. For the first term and for the second term: 
\begin{align*}
    E_{Y|\bm{X}}[\ell_w(\tilde{r} (\bm{X}),Y)|\bm{X}]=&\int_{0}^{1} \{cI_{[\tilde{r}(\bm{X})\ge c]}(1-E_{Y|
\bm{X}}[Y|\bm{X}])+(1-c)I_{[\tilde{r}(\bm{X})\leq c]}E_{Y|
\bm{X}}[Y|\bm{X}]\} w(c)dc \\
=&\int_{0}^{1} \{cI_{[\tilde{r}(\bm{X})\ge c]}(1-\tilde{r}(\bm{X}))+(1-c)I_{[\tilde{r}(\bm{X})\leq c]}\tilde{r}(\bm{X})\} w(c)dc=\ell_w(\tilde{r}(\bm{X}),\tilde{r}(\bm{X})).
\end{align*}
For the second term, additionally, we note that
    $\pi=E_{\bm{X}}[E_{Y|\bm{X}}\{Y|\hat{r}(\bm{X})\}]=E_{\bm{X}}[\tilde{r}(\bm{X})]$
so that $\ell_w(\pi,\pi)=E_{\bm{X}}[\ell_w(\pi,\tilde{r}(\bm{X}))]$. 

\section{Additional simulation results} \label{app:finite}
We repeat the numerical examples in Section 4.1 with a sample size of 800 and 1000 simulation replicates. For each replicate, we construct the 95\% confidence intervals using the bootstrap approach. As showed in Table \ref{tab.app}, the coverage probabilities of the confidence intervals from 1000 replicates are nominal.  

\begin{table}[ht]
\caption{Summary of simulation replicates with a sample size of 800 for the settings in Section 4.1\label{tab.app}}
\centering
\begin{tabular}{rllll@{\hskip 0.3in}llll@{\hskip 0.3in}llll}
 \toprule
 \textbf{Set A} \\

 & \multicolumn{4}{@{}c@{}@{}}{\textbf{Model 1}} & \multicolumn{4}{@{}c@{}@{}}{\textbf{Model 2}} & \multicolumn{4}{@{}c@{}@{}}{\textbf{Model 3}} \\
 \cmidrule{2-13} 
 & Est. & ESE & ASE & CP & Est. & ESE & ASE & CP & Est. & ESE & ASE & CP \\ 
  $BS_w$(1,1) & 0.078 & 0.0035 & 0.0035 & 0.95 & 0.078 & 0.0035 & 0.0035 & 0.95 & 0.089 & 0.0052 & 0.0052 & 0.95 \\ 
  $BS_w$(2,5) & 0.096 & 0.0040 & 0.0039 & 0.94 & 0.073 & 0.0041 & 0.0042 & 0.96 & 0.076 & 0.0049 & 0.0050 & 0.96 \\ 
  $BS_w$(4,8) & 0.110 & 0.0053 & 0.0052 & 0.94 & 0.084 & 0.0049 & 0.0050 & 0.96 & 0.087 & 0.0055 & 0.0056 & 0.96 \\ 
   \midrule
 \textbf{Set B} \\

 & \multicolumn{4}{@{}c@{}@{}}{\textbf{True Model}} & \multicolumn{4}{@{}c@{}@{}}{\textbf{OH Model}} & \multicolumn{4}{@{}c@{}@{}}{\textbf{OL Model}} \\
 \cmidrule{2-13} 
 
 & Est. & ESE & ASE & CP & Est. & ESE & ASE & CP & Est. & ESE & ASE & CP \\ 
  BSw(1,1) & 0.100 & 0.0031 & 0.0032 & 0.96 & 0.107 & 0.0044 & 0.0045 & 0.95 & 0.107 & 0.0045 & 0.0045 & 0.95 \\ 
  BSw(2,5) & 0.107 & 0.0039 & 0.0040 & 0.95 & 0.108 & 0.0041 & 0.0042 & 0.95 & 0.123 & 0.0062 & 0.0062 & 0.95 \\ 
  BSw(4,8) & 0.124 & 0.0048 & 0.0050 & 0.95 & 0.125 & 0.0050 & 0.0052 & 0.95 & 0.141 & 0.0074 & 0.0074 & 0.95 \\ 
   
   \bottomrule
\end{tabular}
\begin{tablenotes}%%[200pt]
\item Est.: average of estimates; ESE: empirical standard error; ASE: average of estimated standard error from the bootstrap; CP: coverage probability of the bootstrap confidence interval; $BS_w(a,b)$: a weighted Brier score with the weight function from $Beta(a,b)$.  
\end{tablenotes}
\end{table}

\section{A weighted Spiegelhalter's Z-statistic} \label{app:wSz}
Within the risk-based decision framework, we may weight Spiegelhalter's Z-statistic to emphasize the assessment of miscalibration in certain risk cutoff regions that are more relevant to clinical decision-making. 

Spiegelhalter's Z-statistic is based on a simple decomposition of the Brier score \citep{rufibach2010use}:
\begin{equation} \label{s_decomp}
    \frac{1}{n} \sum_{i}^{n} (\hat{r}_i-Y_i)^2=\frac{1}{n} \sum_{i}^{n} (Y_i-\hat{r}_i)(1-2\hat{r}_i)+\frac{1}{n} \sum_{i}^{n} \hat{r}_i(1-\hat{r}_i).
\end{equation}
 
Under the well-calibration assumption (i.e., $P(Y_i|\hat{r}_i)=\hat{r}_i$), the first summand has expectation 0 conditional on $\hat{r}_i$, and we have an extra term: $E_0[\frac{1}{n} \sum_{i}^{n} (\hat{r}_i-Y_i)^2|\hat{r}_i]=\frac{1}{n} \sum_{i}^{n} \hat{r}_i(1-\hat{r}_i)$. To exploit this fact, \cite{spiegelhalter1986probabilistic} constructed a Z-statistic that is approximately standard normal under the well-calibration assumption:

\begin{equation} \label{Spiegelhalter}
\begin{split}
        Z_{Spiegelhalter}& =\frac{1/n \sum_{i}^{n} (\hat{r}_i-Y_i)^2-E[1/n \sum_{i}^{n} (\hat{r}_i-Y_i)^2|\hat{r}_i]}{\sqrt{Var_0[1/n \sum_{i}^{n} (\hat{r}_i-Y_i)^2|\hat{r}_i]}} \\
        &=\frac{\sum_{i}^{n} (Y_i-\hat{r}_i)(1-2\hat{r}_i)}{\sqrt{\sum_{i}^{n} (1-2\hat{r}_i)^2\hat{r}_i(1-\hat{r}_i)}},
\end{split}
\end{equation}
where $Var_0$ is the variance under the well-calibration assumption.  

Spiegelhalter's Z-statistic has been implemented in STATA \citep{Stata_statabase} and a popular r-package \textit{rms} \citep{harrell2015regression}, and has some applications in the risk prediction research to assess calibration \citep[e.g., ][]{walsh2017beyond}. \cite{huang2020tutorial} suggested that Spiegelhalter's Z-statistic is not intuitive; thus, it is not widely used in practice compared to the Hosmer–Lemeshow statistic and the calibration plot. It is intuitive to assess the regions of miscalibration on the risk spectrum from the calibration plot or each summand in the Hosmer–Lemeshow statistic. On the contrary, Spiegelhalter's Z-statistic assesses the calibration in a global sense, which is inherited from the Brier score as an overall summary. To alleviate this shortcoming, we can construct a weighted Spiegelhalter's Z-statistic stemming from the weighted Brier score. 
We can adopt the notations from Section \ref{sect decomp} to rewrite a general decomposition of (\ref{s_decomp}) as: 

\begin{equation} \label{ws_decomp}
    \begin{split}
        \frac{1}{n}\sum_{i=1}^n[\ell_w(\hat{r}_i,Y_i)]= & (\frac{1}{n}\sum_{i=1}^n[\ell_w(\hat{r}_i,Y_i)]-\frac{1}{n}\sum_{i=1}^n[\ell_w(\hat{r}_i,\hat{r}_i)])+\frac{1}{n}\sum_{i=1}^n[\ell_w(\hat{r}_i,\hat{r}_i)] \\
        = & \frac{1}{n}\sum_{i=1}^n[(Y_i-\hat{r}_i)(1-F_w(\hat{r}_i)-\mu_w)]+\frac{1}{n}\sum_{i=1}^n[\ell_w(\hat{r}_i,\hat{r}_i)].
    \end{split}
\end{equation}

Following the similar construction for Spiegelhalter's Z-statistic, we can define a weighted version: 

\begin{equation} \label{wSz}
   \begin{split}
    wZ_{Spiegelhalter}=\frac{\frac{1}{n}\sum_{i=1}^n[\ell_w(\hat{r}_i,Y_i)]-\frac{1}{n}\sum_{i=1}^n[\ell_w(\hat{r}_i,\hat{r}_i)]}{\sigma_{0n}/\sqrt{n}} =\frac{\frac{1}{n}\sum_{i=1}^n[(Y_i-\hat{r}_i)(1-F_w(\hat{r}_i)-\mu_w)]}{\sigma_{0n}/\sqrt{n}},
    \end{split}
\end{equation}
where $\sigma_{0n}$ is the square root of the estimated variance of $\ell_w(\hat{r}_i,Y_i)$ under the null hypothesis of well-calibration (see Appendix for details). It is easy to verify that when $w(c)$ is the uniform distribution over (0,1), Equations (\ref{s_decomp}) and (\ref{Spiegelhalter}) are special cases of (\ref{ws_decomp}) and (\ref{wSz}), respectively.

%The weighted Spiegelhalter's Z-statistic might still be less intuitive for the practitioners and more abstract than the calibration plot. It is beyond the scope of this paper to compare which means of calibration assessment are preferred; thus, we do not forcefully promote the weighted Spiegelhalter's Z-statistic over the other means. 

\begin{remark}

The purpose of the discussion on the weighted Spiegelhalter's Z-statistic as well as the $H$ measure is to show how ideas of the weighted Brier score can be interesting to researchers who focus on methodological development. For the weighted Brier score, the wide adoption of the $H$ measure in the machine learning community additionally validates the proposed weighting method. The lack of calibration in the $H$ measure coincides with over-emphasis on the AUC and relatively little attention to calibration, which is referred to as `the Achilles heel of predictive analytics' \citep {van2019calibration}.  On the other hand, some researchers may simply equate the Brier score with calibration, which is incorrect because of the extra term in the decomposition of Spiegelhalter's Z-statistic \citep{rufibach2010use}. The swift construction of the weighted Spiegelhalter’s Z-statistic highlights the promising possibilities for novel methods that emerge from a deeper understanding of the Brier score. Indeed, as researchers who are interested in methodological development, we appreciate the elegant and practically meaningful mathematical structures that lie beneath the deceptively simple and familiar Brier score. 
\end{remark}

\end{document}